\documentclass[conference,10pt]{IEEEtran}
\IEEEoverridecommandlockouts
\usepackage{cite}
\usepackage{amsmath,amssymb,amsfonts}
\usepackage{algorithmic}
\usepackage{graphicx}
\usepackage{textcomp}
\usepackage{xcolor}
\usepackage{enumerate}
\usepackage{pifont}
\usepackage{romannum}
\usepackage{multirow}

\usepackage{tabularx}

\newcolumntype{L}[1]{>{\ce\raggedright\arraybackslash}p{#1}}
\newcolumntype{C}[1]{>{\centering\arraybackslash}p{#1}}
\newcolumntype{R}[1]{>{\raggedleft\arraybackslash}p{#1}}

\usepackage[left=0.67in, right=0.67in, top=0.60in, bottom=0.65in]{geometry}

\usepackage{setspace}
\newcommand{\ignore}[1]{}
\def\BibTeX{{\rm B\kern-.05em{\sc i\kern-.025em b}\kern-.08em
    T\kern-.1667em\lower.7ex\hbox{E}\kern-.125emX}}
\begin{document}

\bstctlcite{IEEEexample:BSTcontrol}

\title{TaxoNN: A Light-Weight Accelerator for\\ Deep Neural Network Training}


\author{Reza Hojabr\IEEEauthorrefmark{2}\IEEEauthorrefmark{4}, Kamyar Givaki\IEEEauthorrefmark{2}$^{,1}$, Kossar Pourahmadi\IEEEauthorrefmark{2}$^{,1}$, Parsa  Nooralinejad\IEEEauthorrefmark{2}$^{,1}$,\\Ahmad Khonsari\IEEEauthorrefmark{2}\IEEEauthorrefmark{3}, Dara Rahmati\IEEEauthorrefmark{3}, M. Hassan Najafi\IEEEauthorrefmark{4}\\
\IEEEauthorrefmark{2}School of Electrical and Computer Engineering, University of Tehran, Tehran, Iran \\
\IEEEauthorrefmark{3}School of Computer Sciences, Institute for Research in Fundamental Sciences (IPM), Tehran, Iran\\
\IEEEauthorrefmark{4}School of Computing and Informatics, University of Louisiana at Lafayette, Lafayette, LA, USA\\
\{r.hojabr,givakik,kosar.pourahmadi,p.nooralinejad\}@ut.ac.ir, \{ak,dara.rahmati\}@ipm.ir, najafi@louisiana.edu
\vspace{-0.5em}
}

\maketitle

\begin{abstract}
Emerging intelligent embedded devices rely on Deep Neural Networks (DNNs) to be able to interact with the real-world environment. This interaction comes with the ability to retrain DNNs, since environmental conditions change continuously in time. Stochastic Gradient Descent (SGD) is a widely used algorithm to train DNNs by optimizing the parameters over the training data iteratively.
In this work, first we present a novel approach to add the training ability to a baseline DNN accelerator (inference only) by splitting the SGD algorithm into simple computational elements. Then, based on this heuristic approach we propose \textit{TaxoNN, a light-weight accelerator for DNN training}. TaxoNN can easily tune the DNN weights by reusing the hardware resources used in the inference process using a time-multiplexing approach and low-bitwidth units.
Our experimental results show that TaxoNN delivers, on average,  0.97\% higher misclassification rate compared to a full-precision implementation. Moreover, TaxoNN provides 2.1$\times$ power saving and 1.65$\times$ area reduction over the state-of-the-art DNN training accelerator.

\end{abstract}


\section{Introduction}
Driven by the availability of large datasets, deep learning applications are increasingly growing in various fields such as speech recognition, computer vision, control and robotics. Meanwhile, time-consuming computations of DNNs and the need for power-efficient hardware implementations have made the semiconductor industry to rethink the customized hardware for deep learning algorithms. As a result, DNN hardware accelerators have been emerged as a promising solution to tackle efficient implementation of these compute-intensive and energy-hungry algorithms\cite{eyeriss,diannao,shidiannao}.\color{white}\footnote{\vspace{-2.5em}~These authors contributed equally.}\color{black}

Employing deep learning algorithms in building intelligent embedded devices that interact with the environment requires customized accelerators that support both training and inference processes. For instance, in deep reinforcement learning algorithms, an agent uses a neural network (NN) to predict the proper action regarding the current state and the reward obtained from the environment. In such algorithms that a NN-based agent is interacting with the  environment and the environmental conditions are changing continuously, the training process is performed repeatedly to tune the agent. Implementing the prohibitive computations of the training process seeks an efficient yet low-power trainable architecture. Although the recently proposed accelerators have significantly improved the performance of the inference process~\cite{eyeriss}, there is still a growing demand for low-power DNN accelerators that support both training and inference processes.

Training process can be interpreted as an optimization problem that aims to minimize an objective function (network error function) by finding a set of network parameters. Stochastic Gradient Descent (SGD) is a common approach to solve this optimization problem \cite{dorefanet}. SGD moves towards the optimum point in the decreasing direction of the error function's gradient.
Calculating the gradients during SGD requires high-cost hardware resources which cannot be provided in embedded devices with limited power and area budget. Therefore, adding the training capability to the conventional inference-only accelerators is a challenging issue and needs a complete rethink. This work seeks a solution to enable inference-only accelerators to perform SGD computations with minimum hardware resources.

Relying on the approximate nature of NNs, several methods have been proposed replacing floating point units of NN with low-bitwidth ones\cite{bitfusion,CORN,stripes,cambricon}. Prior work have shown that employing low-bitwidth operations in DNN accelerators can result in a substantial power and area saving while maintaining the quality of the results~\cite{stripes,bitfusion,dorefanet}. While it is more common to employ low-bitwidth data in the inference process, recent works have demonstrated that the training process (i.e., SGD) can also be performed using quantized parameters~\cite{dorefanet}. Our observations  confirm that the desired accuracy can be achieved, without sacrificing the network convergence, when using low-bitwidth operations during the training process. An important point, however, is that the required bitwidth can vary from layer to layer. As we get closer to final layers of DNNs, the extracted features become more valuable. While the early layers produce satisfying results with small bitwidths, a more precise computation is necessary in the final layers. Leveraging this observation, by proper adjustment of the bitwidth in each layer we can reduce a significant amount of power and area while  maintaining the quality of the results. 

This work proposes a novel low-cost accelerator that supports both training and inference processes. We first propose a novel method to split the SGD algorithm into smaller computational elements by unrolling this compute-intensive algorithm. Using our proposed method, a fine-grained inter-layer parallelism can be used in the training process. We then leverage this method and introduce TaxoNN, \textit{a Light-Weight Accelerator for DNN training} which is able to perform training and inference processes using shared resources. TaxoNN utilizes an optimized datapath in a pipelined manner that minimizes the hardware cost. We show how bitwidth optimization in different layers of NN can reduce the implementation cost while keeping the quality of the results. In summary, the main contributions of this work are as follows:

\begin{enumerate}[i.]
    \item We propose a novel heuristic method to minimize the implementation cost of the SGD algorithm by unrolling its computations. The proposed method reduces the hardware cost by time-division multiplexing (TDM) of multiply-and-accumulate (MAC) units.
    
	\item We introduce an accelerator for DNN training, called TaxoNN, that supports training and inference using this method. TaxoNN parallelizes the SGD algorithm while minimizes the required arithmetic units.
	
	\item We evaluate TaxoNN in terms of network convergence, power consumption, and area using low-bitwidth computational units for different layers. Our experimental results show that TaxoNN offers 2.1$\times$ power and 1.65$\times$ area saving at the cost of\color{black}, on average, 0.97\% higher misclassification rate compared to the full-precision implementation. 
\end{enumerate}

\section{Related Work}
Plenty of work have introduced specialized accelerators for deep learning~\cite{shidiannao,minerva,diannao}. Motivated by the processing characteristics of DNNs, Eyeriss \cite{eyeriss} introduced a novel data-flow to maximize data reuse between neural Processing Elements (PEs) and hence to minimize the energy consumption wasted on data movements. As there are various types of layers in DNNs (convolutional, pooling and fully connected), MAERI \cite{maeri} and FlexFlow \cite{flexflow} proposed new design methodologies to enable flexible data-flow mapping over neural accelerators. 
Moreover, eliminating unnecessary multiplications in sparse layers \cite{scnn,cambricon,cnvlutin} and computation reuse \cite{CORN,computationreuse,ucnn} are promising solutions to reduce the cost of DNN accelerators.

Recently, replacing full-precision operations with low-bitwidth ones has been used as an effective approach to save energy consumption of DNNs~\cite{dynamic,bitfusion, skippynn}. Experimental observations have shown that the approximate nature of DNNs makes them tolerable to the quantization noise~\cite{CORN,bitfusion,approximate}. Hence, costly floating-point arithmetic units are replaced by fixed-point ones at no considerable accuracy loss.  Bit Fusion~\cite{bitfusion} presents a bit-level flexible accelerator that dynamically sets the bitwidth to minimize the computation cost. 

While the focus of most prior work has been on developing high-performance architectures for the inference process, some recent work proposed accelerators for training DNNs \cite{pipelayer,time,tnpu}. TIME \cite{time} and Pipelayer \cite{pipelayer} utilized Process-In-Memory (PIM) techniques to accelerate the training process. Performing the operations near memory helps to alleviate the data movement overhead during the DNN computations. However, to the best of our knowledge, no hardware architecture and datapath have been developed to reduce the processing time of the SGD algorithm by exploiting parallelism in its heavy computations.
Some recent work have also shown that training can be performed using low-bitwidth gradients \cite{dorefanet,quantized,wage,flexpoint}. In this work, we minimize the overall cost of the proposed accelerator, TaxoNN, by proper adjustment of bitwidth in each layer of the network.
\begin{figure}[!t]
	\centering
	\includegraphics[width=0.47\textwidth]{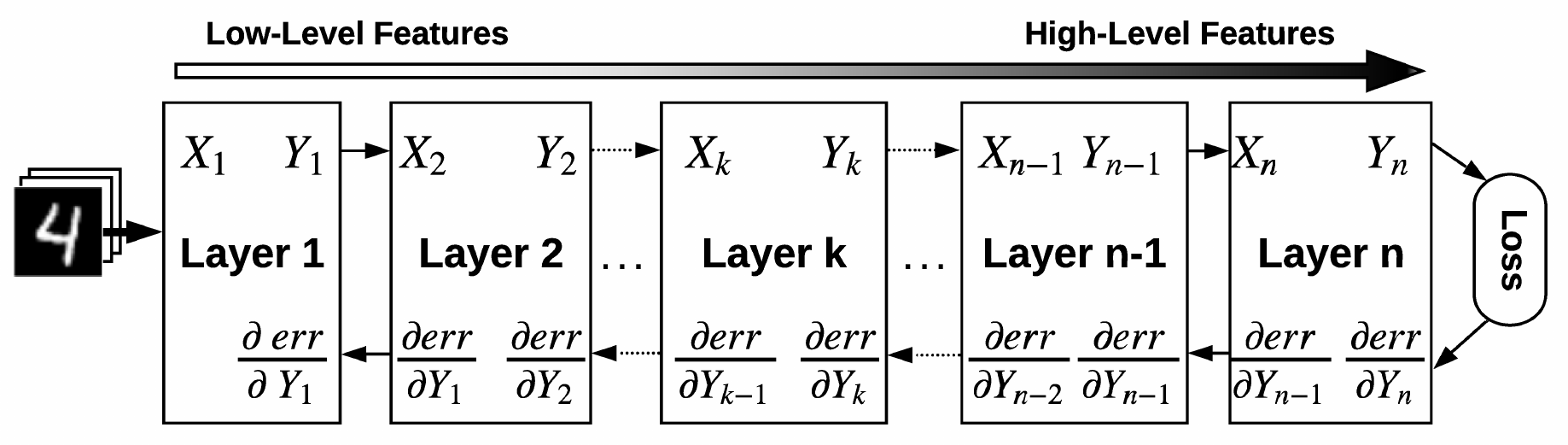}
	\vspace{-.5em}
	\centering
	\caption{Back-propagation in the layers of a DNN.}
	\label{fig1}
	\centering
	\vspace{-1em}
\end{figure}

\section{Motivation and Background}
The training process has the most prominent role in designing an accurate DNN. The underlying principle in training methods arises from what occurs in the human brain. To distinguish a certain object, a set of various pictures demonstrating the object in different gestures are fed to the network in an iterative manner. The network gradually learns to identify an object  by extracting its features in multiple iterations. By comparing the output to the desired result, the network learns how to change the parameters. This procedure continues until the network finds the best weights that maximize the recognition accuracy.

From mathematical point of view, training procedure is performed by an error Back-Propagation (BP) method. As depicted in Fig. \ref{fig1}, an input data is fed to the network and forwarded through the layers. The produced output is fed to a loss function to calculate the gradient of the error. The computed gradient is then back-propagated through the layers to update the weights. During back-propagation, the gradient of the error tends gradually to zero. This method is called Gradient Descent. Eq.~\ref{eq:1} shows how the weights in layer $i$ are updated by the gradient. Learning rate is shown with $\alpha$ which determines the rate of network convergence by controlling the impact of gradients during the training process.  Due to large amount of data, feeding all inputs to the network is very time-consuming. Therefore, a subset of data is picked up randomly in each iteration to train the network. This method, called Stochastic Gradient Descent (SGD), is the most common approach to train DNNs. 
{\footnotesize
\begin{equation} 
	\label{eq:1}
	W_{i} = W_{i} - \alpha  \frac{\partial error}{\partial{W_{i}}}
\end{equation}
}
\normalsize
Training often takes a long time to be completed as its processing time is directly proportional to the number of layers. Conventional DNNs are composed of a large number of layers (may even more than a thousand layers). Convolutional layers constitute the most portion of the computation load in DNNs. These layers are obligated to extract the features of the input data. Normally, the early layers extract general features that can be used in distinguishing any object. As we get closer to final layers, we extract more valuable features that help to recognize specific objects.
\section{Proposed Architecture}
Due to limited processing resources in embedded systems, TaxoNN aims to train the network by reusing hardware resources used in inference to minimize the hardware cost. In what follows, we describe the micro-architecture of TaxoNN.
\subsection{Inference Architecture}
\begin{figure}[!t]
	\centering
	\includegraphics[width=0.47\textwidth]{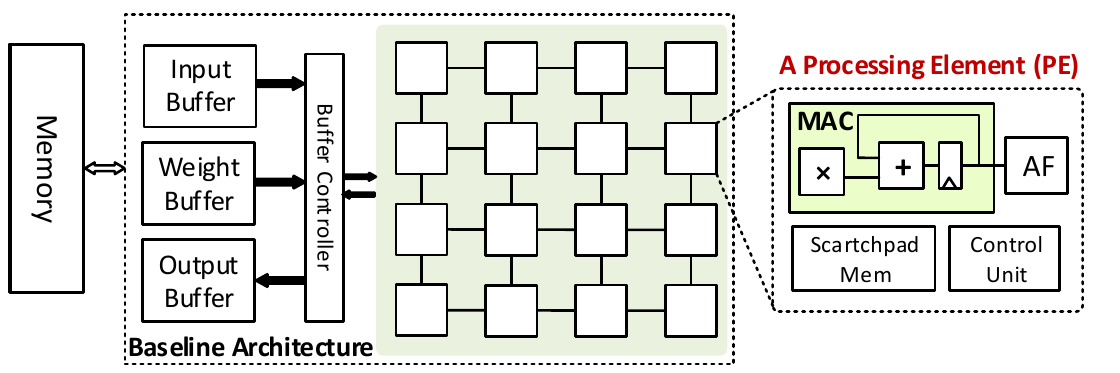}
	\vspace{-0.75em}
	\centering
	\caption{The baseline architecture of TaxoNN to execute the inference of DNNs}
	\label{inference}
	\centering
	\vspace{-1em}
\end{figure}

The baseline architecture of TaxoNN, designed to perform the inference process, is shown in Fig.~\ref{inference}. Similar to the state-of-the-art  accelerator~\cite{eyeriss}, it is composed of a 2D array of Processing Elements (PEs) used in both convolutional and fully-connected (FC) layers.  In general, the output of each neuron (a.k.a filter in the convolutional layers) is achieved by a weighted summation, $y=f(\sum_{i=0}^{i=k} x_{i}w_{i})$, where $x_{i}$ is the input vector, $w_{i}$ is the weight vector and $f$ denotes the activation function. The activation function is typically Sigmoid in FC layers and ReLU (Rectified Linear Unit) in convolutional layers. In TaxoNN, each layer is equipped with input/output buffers that fetch/store the input/output data. Each PE can access to the weight buffer that holds the weight vector.
To decrease the number of data accesses to the input buffer, the fetched values are forwarded through the PEs in a pipelined manner. PEs are equipped with a local scratchpad memory to hold the weights and partial results.

In the FC layer $L_{i}$, the required time to complete the computations of the neurons is $N_{i-1}+N_{i}$ clock cycles where $N_{i-1}$ and $N_{i}$ are the number of neurons in the $(i-1)^{th}$ and $i^{th}$ layers, respectively (provided that we have $N_i$ PEs). In the convolutional layer $L_{j}$, where the input data size is $h \times w \times d$ and the filter size is $k \times k \times d$, the convolution is achieved in $kd(w-k)(h-k)$ clock cycles. Similar to the \textit{Row-Stationary} dataflow proposed in Eyeriss~\cite{eyeriss}, each compute lane in the PE array is dedicated to a single row of the filter to maximize the data reuse in the architecture.
\subsection{Simplifying SGD algorithm}
\label{training-process}
As mentioned in Eq. (\ref{eq:1}), weights are  updated in each layer by subtracting the term $\alpha \frac{\partial error}{\partial{W_{i}}}$ from their current value. The first step towards enabling training in the porposed accelerator is to simplify the term $ \frac{\partial error}{\partial{W_{i}}}$ to implement it with the minimum hardware resources. Leveraging the chain rule we can partition $ \frac{\partial error}{\partial{W_{i}}}$ into three small parts as follows:

{\footnotesize
\begin{equation} \label{eq:3}
	\frac{\partial error}{\partial{W_{i}}}= \frac{\partial error}{\partial{Y_{i+1}}}\times  \frac{\partial Y_{i+1}}{\partial{Y_{i}}}\times \frac{\partial Y_{i}}{\partial{W_{i}}}
\end{equation}
}\normalsize
where $Y_{i}$ is the output of the $i^{th}$ layer. Note that all the notations are written in the matrix form. The terms $\frac{\partial Y_{i+1}}{\partial{Y_{i}}}$  and $\frac{\partial Y_{i}}{\partial{W_{i}}}$  can further be expanded as follows:
{\footnotesize
\begin{equation} \label{eq:4}
	\frac{\partial Y_{i+1}}{\partial{Y_{i}}}=\frac{\partial f_{i+1}(W_{i+1}Y_{i})}{\partial{Y_{i}}}=W_{i+1}{f}'_{i+1}(W_{i+1}Y_{i})
\end{equation}
\begin{equation} \label{eq:5}
	\frac{\partial Y_{i}}{\partial{W_{i}}}=\frac{\partial f_{i}(W_{i}Y_{i-1})}{\partial{W_{i}}}=Y_{i-1}{f}'_{i}(W_{i}Y_{i-1})
\end{equation}
}\normalsize
where $f_{i+1}(.)$ denotes the activation function of the $(i+1)^{th}$ layer and ${f}'$ refers to the derivation of the activation function. Combining Eq.~(\ref{eq:4}) and Eq.~(\ref{eq:5}) leads to Eq.~(\ref{eq:6}):
{\footnotesize
\begin{equation} \label{eq:6}
	\frac{\partial error}{\partial{W_{i}}}= \frac{\partial error}{\partial{Y_{i+1}}}\times f'_{i+1} \times W_{i+1} \times f'_{i} \times Y_{i-1}
\end{equation}
}
\normalsize
We define $G_{i+1}$  as the product of the first two terms in the right hand side (RHS) of Eq. (\ref{eq:6}) that is computed in the $(i+1)^{th}$ layer and passed backward to the $i^{th}$ layer. 
{\footnotesize
\begin{equation} \label{eq:10}
	G_{i+1} = \frac{\partial error}{\partial{Y_{i+1}}} \times {f}'_{i+1}
\end{equation}
}
\normalsize
Clearly,  the input of $i^{th}$ layer is the output of $(i-1)^{th}$ layer ($X_{i}=Y_{i-1}$, where $X_{i}$ is the input of $i^{th}$ layer). As a result, we can rewrite Eq.~(\ref{eq:3}) as follows:
{\footnotesize
\begin{equation} \label{eq:7}
	\frac{\partial error}{\partial{W_{i}}}= G_{i+1} \times W_{i+1} \times f'_{i} \times X_{i}
\end{equation}
}
\normalsize
To facilitate the hardware implementation, Eq.~(\ref{eq:7}) can be split into Eq.~(\ref{eq:8}) and Eq.~(\ref{eq:9}). As shown in Eq.~(\ref{eq:9}), multiplying Eq.~(\ref{eq:8}) by the input of $i^{th}$layer, $X_{i}$, results in term $\frac{\partial error}{\partial{W_{i}}}$ in Eq. (\ref{eq:1}). 
{\footnotesize
\begin{equation} \label{eq:8}
	G_{i}= G_{i+1} \times W_{i+1} \times f'_{i}
\end{equation}
\begin{equation} \label{eq:9}
	\frac{\partial error}{\partial{W_{i}}}= G_{i} \times X_{i}
\end{equation}
}
\normalsize

Consequently, $G_{i}$ has a key role in the training process. As shown in Eq. (\ref{eq:8}), $G_{i}$ is achieved recursively by calculating in each layer and passing backward to the previous layer. We use this unrolling method to distinguish between the operations in the SGD and to properly map them to the hardware resources.\color{black}

\begin{figure}[h]
	\vspace{-.5em}
	\centering
	\includegraphics[width=0.47\textwidth]{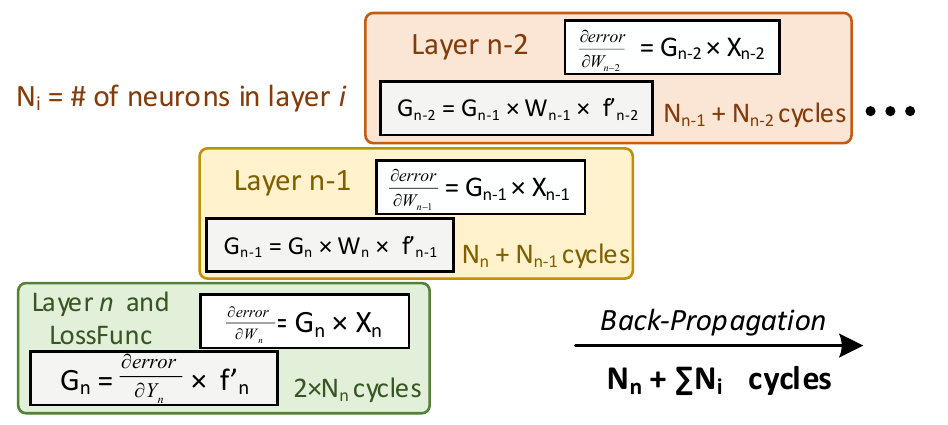}
	\centering
	\vspace{-1em}
	\caption{Timing diagram of the training process in TaxoNN.}
	\label{timing}
	\centering
	\vspace{-1.em}
\end{figure}

\begin{figure}[!t]
	\centering
	\includegraphics[width=0.49\textwidth]{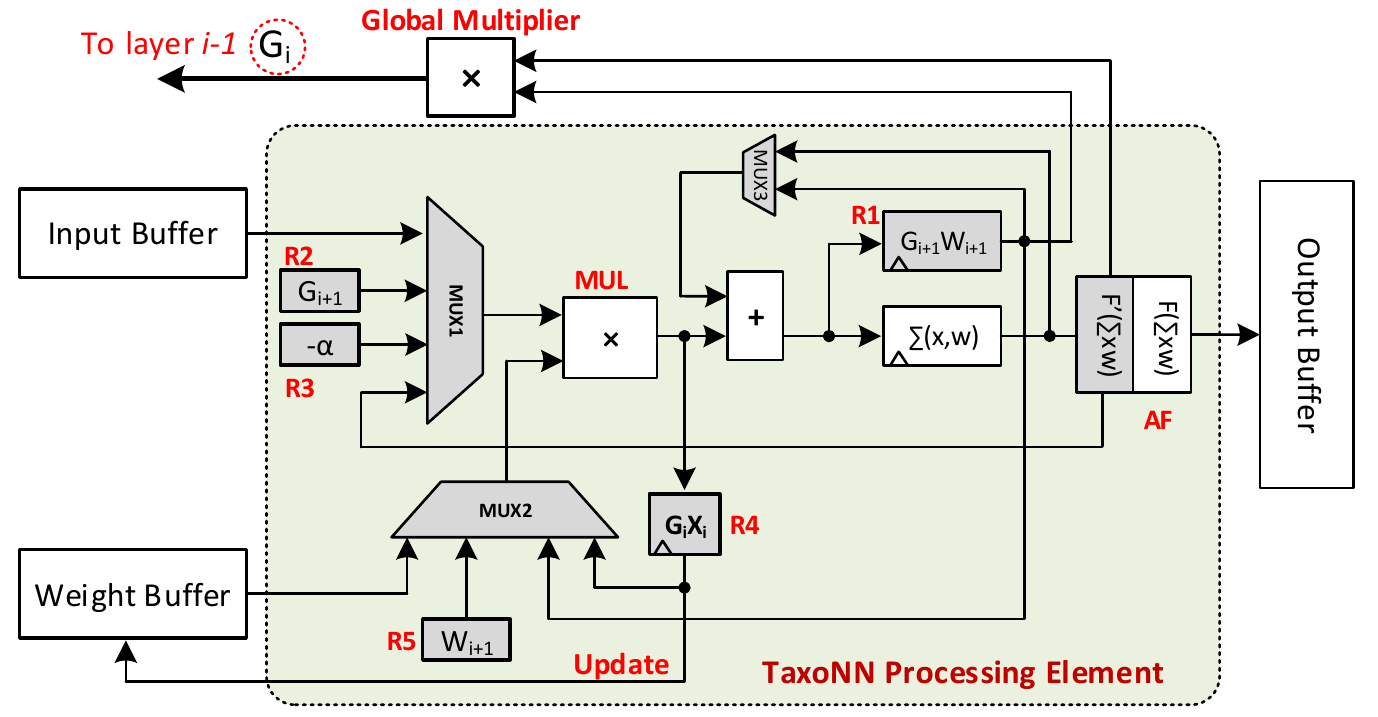}
	\centering
	\vspace{-1.5em}
	\caption{The micro-architecture of a PE in TaxoNN with the training capability.}
	\label{TaxoNN}
	\centering
	\vspace{-1.em}
\end{figure}

\subsection{Training Architecture}
To implement the BP computations, the baseline architecture must be modified by adding some simple logical components. Fig.~\ref{TaxoNN} illustrates the micro-architecture of the proposed PEs in TaxoNN. The gray components are added to the baseline architecture to enable training. To minimize the needed resources, we employ a TDM approach to improve the resource utilization of the main components (e.g., the multipliers) in the datapath. In what follows, we describe each component in detail.

\noindent\textbf{Multiplexers.} As depicted in Fig.~\ref{TaxoNN}, the architecture is equipped with three multiplexers to enable TDM.  The inference process is still performed using the main multiplier. All the needed parameters of Eq.~(\ref{eq:8}) and Eq.~(\ref{eq:9}) can be provided by a proper timing management of MUX1, MUX2 and the multiplier as follows:
\ding{182} MUX1 provides $G_{i+1}$ and MUX2 provides $W_{i+1}$. Then, $G_{i+1} \times W_{i+1}$ will be calculated and stored in register R1. 
\ding{183}~MUX1 forwards ${f}'_{i}$ and MUX2 forwards $G_{i+1} \times W_{i+1}$ to the multiplier to calculate $G_{i}=G_{i+1} \times W_{i+1} \times {f}'_{i}$.
\ding{184} MUX1 forwards $X_{i}$ to multiply it by $G_{i}$ and hence produce $\frac{\partial error}{\partial{W_{i}}}$.
\ding{185}~Finally, the result is multiplied by the learning rate, $\alpha$, that is already stored in a register behind MUX1.

In this manner, $-\alpha \frac{\partial error}{\partial{W_{i}}}$ as the most important parameter for updating the weights, is prepared through a TDM of the PE's multiplier. Note that $G_{i+1}$ and $W_{i+1}$ have been provided and sent to the current layer by the $(i+1)^{th}$ layer. Since all the computations are done in the matrix form, calculating $G_{i+1} \times W_{i+1}$ needs $N_{i+1}$ cycles where $N_{i+1}$ is the number of neurons in the $(i+1)^{th}$ layer. After each multiplication, the result is accumulated in the corresponding register. 

\noindent\textbf{Activation Function.} The activation function unit of the baseline architecture (Fig.~\ref{inference}) is equipped with an internal unit to calculate the derivation of the activation functions. There are three types of activation functions which are commonly used in the modern DNNs: ReLU, Sigmoid and $tanh$. The derivation of  sigmoid $\sigma(x)$ can be easily achieved by ${\sigma}'(x)=\sigma(x)(1-\sigma(x))$. Also, $tanh$ is simply achieved from $\sigma(x)$ as $tanh(x)=2\sigma (2x)-1$ and consequently, ${tanh}'(x)$ can be achieved as ${tanh}'(x)=4{\sigma}'(2x)$. Finally, the derivation of the ReLU is 0 for negative inputs and 1 for positive ones.

\noindent\textbf{Global Multiplier.}
In TaxoNN, each layer $i$ has a single global multiplier to produce $G_{i}$. This multiplier is shared between all the neurons of the layer. Therefore, the number of cycles needed to produce $G_{i}$ equals the number of neurons in that layer.

Consequently, the following components are added to the baseline PE: \textit{(\romannum{1})} three multiplexers, \textit{(\romannum{2})} five registers (located in the scratchpad memory to hold the intermediate values during training), and \textit{(\romannum{3})} activation function's derivation unit. The overhead cost of these components will be discussed in Section~\ref{performance-evaluation}.
\begin{figure*}[!h]
	\centering
	\includegraphics[width=0.9\textwidth]{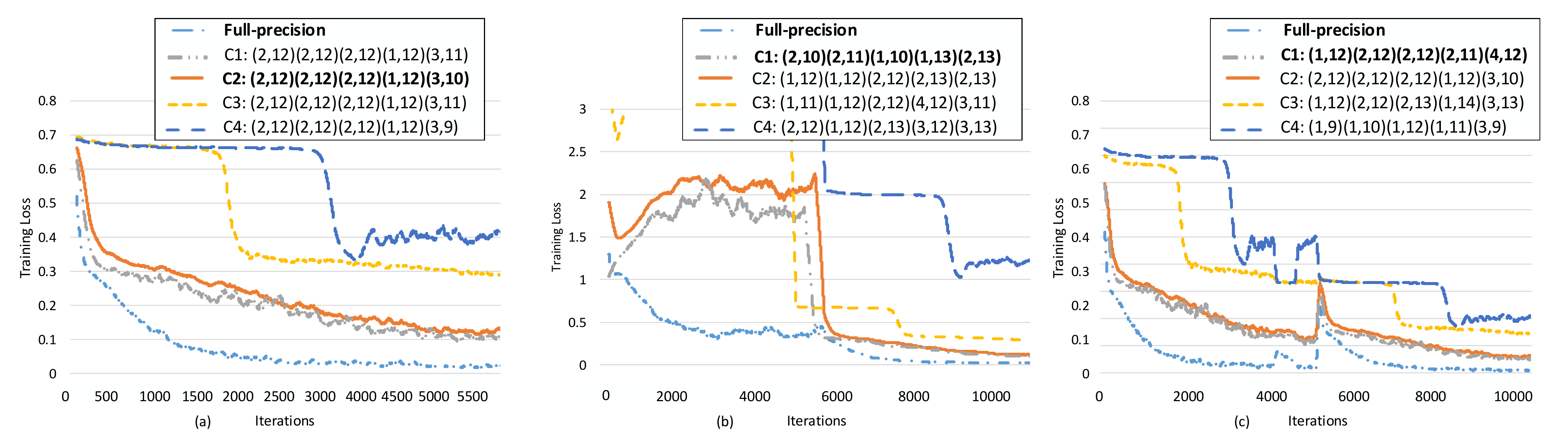}
	\centering
	\vspace{-1.25em}
	\caption{The network loss: (a) MNIST and (b) CIFAR10, (c) SVHN.}
	\label{loss}
	\centering
\vspace{-1.5em}
\end{figure*}

\subsection{Timing and Pipeline}
TaxoNN benefits from an optimized and pipelined architecture. Fig.~\ref{timing} shows the timing diagram of the training process composed of the forwarding phase followed by the error BP and weight updating. As mentioned in Section~\ref{training-process}, $G_{i}$ is the main precedence for calculating $-\alpha \frac{\partial error}{\partial{W_{i}}}$. In layer $i$, $G_{i}$ is a vector of size $N_{i}$, where $N_{i}$ is the number of neurons in that layer. Whenever $G_{i,1}$ (the first element of the matrix $G_{i}$) becomes ready, it will be sent to the previous layer, $L_{i-1}$, that needs the elements of  $G_{i}$ to calculate  $ \frac{\partial error}{\partial{W_{i-1}}}$. Therefore, producing $ \frac{\partial error}{\partial{W_{i}}}$ has a timing overlap with producing $G_{i}$ in the $(i)^{th}$ layer. Leveraging this pipelining, TaxoNN performs an iteration of the BP in $N_{n}+\sum_{i=1}^{i=n} N_{i}$ clock cycles, where $n$ is the number of layers. The extra $N_{n}$ is for the computations of the loss function and is equal to the processing time of the last layer ($n^{th}$ layer).

\section{Performance Evaluation}
\label{performance-evaluation}
We used the LeNet architecture~\cite{lenet} to evaluate the performance of TaxoNN using MNIST, CIFAR10 and SVHN datasets. We extracted the results of full-precision computations using TensorFlow. In what follows, we analyze TaxoNN in terms of accuracy, network convergence, and hardware cost.

\subsection{Bitwidth Optimization}
Fig.~\ref{loss} demonstrates the network loss during different iterations of the training process using TaxoNN with different bitwidths (optimized for each layer) versus the case of training using the full-precision implementation. MNIST and SVHN are two datasets consist of $28 \times 28$  images from hand-written digits~(0..9) and house numbers, respectively. CIFAR10 is a set of $32 \times 32$ color images in 10 classes.  The training performance is evaluated over 10,000 test images and the network accuracies are extracted by TensorFlow.

The results shown in Fig.~\ref{loss} confirm  that the low-bitwidth training can have \color{black} a comparable accuracy for the same number of iterations. 
The optimum bitwidth for each layer can be different from other layers. For each dataset, we evaluated the network accuracy for a large number of design points. Fig.~\ref{loss} shows four design points for each dataset, each point representing the adopted precision for a layer. The number representation ($I$,$F$) indicates a fixed-point number with $I$ bits for the integer part and $F$ bits for the fractional part.

For instance, during the training of MNIST, the configuraction set C2 converges similar to the floating-point implementation. \color{black}Lower bitwidths, however, may cause under-fitting. The speed of the network convergence gets reduced as the bitwidth gets shorter. This phenomenon  implies that the network confidence is directly related to the precision of the arithmetic operations. An observation is that there is a lower bound that limits the bitwidth of the training. The bitwidths lower than these thresholds cause under-fitting while the bitwidths higher than them are not necessary and will only cost additional  area and power consumption.
\color{black}
\begin{table}[!t]
\centering
\caption{The network accuracy (\%) of different bitwidth versus the floating-point implementation.}
\vspace{-0.75em}
\label{summary}
\renewcommand{\arraystretch}{0.9}
\resizebox{3.2in}{!}
{
\begin{tabular}{|c|c|c|c|}
\hline
Dataset & Precision per Layer (I,F)      & \begin{tabular}[c]{@{}c@{}}TaxoNN \\ Accuracy\end{tabular} & \begin{tabular}[c]{@{}c@{}}Full-precision \\ Accuracy\end{tabular} \\ \hline
MNIST   & (2,12)(2,12)(2,12)(1,12)(3,10) & 99.1                                                       & 99.4                                                               \\ \hline
CIFAR10 & (2,10)(2,11)(1,10)(1,13)(2,13) & 84.1                                                       & 85.4                                                               \\ \hline
SVHN    & (1,12)(2,12)(2,12)(2,11)(4,12) & 94.7                                                       & 96.0                                                                 \\ \hline
\end{tabular}
}
\vspace{-1em}
\end{table}

\begin{table}[t]
\centering
\caption{The area $(um^{2}\times 10^3)$ of a processing element of  TaxoNN versus that of the baseline architecture.}
\vspace{-0.75em}
\renewcommand{\arraystretch}{1.0}
\setlength{\tabcolsep}{1.1pt}
\resizebox{3.4in}{!}
{
\label{area}
\begin{tabular}{|c|c|c|c|c|c|c|c|c|c|c|}
\hline
Bitwidth & 21 & 20 & 19 & 18 & 17 & 16 & 15 & 14 & 13 & \multirow{3}{*}{\begin{tabular}[c]{@{}c@{}}Average\\ Area\\ Overhead\end{tabular}} \\ \cline{1-10}
Eyeriss & 14.3  & 13.1  & 11.8  & 11.1  & 10.6  & 10.1  & 9.7   & 9.0   & 8.1   &                                                                                   \\ \cline{1-10}
\textbf{TaxoNN}   & 15.5  & 14.3  & 12.9  & 12.1  & 11.7  & 11.2  & 10.6  & 9.9   & 9.0   &                                                                                   \\ \hline
Overhead & 8.3\% & 9.2\% & 9.1\% & 8.6\% & 10.0\% & 10.8\% & 8.8\% & 9.8\% & 10.5\% & 9.5\%                                                                            \\ \hline
\end{tabular}
}
\vspace{-1em}
\end{table}

\begin{table}[t]
\centering
\caption{The power consumption $(mW)$ of a processing element of TaxoNN versus that of the baseline architecture.}
\vspace{-0.75em}
\label{power}
\setlength{\tabcolsep}{1.1pt}
\renewcommand{\arraystretch}{1.0}
\resizebox{3.5in}{!}
{
\begin{tabular}{|c|c|c|c|c|c|c|c|c|c|c|}
\hline
Bitwidth & 21 & 20 & 19 & 18 & 17 & 16 & 15 & 14 & 13 & \multirow{3}{*}{\begin{tabular}[c]{@{}c@{}}Average\\ Power\\ Overhead\end{tabular}} \\ \cline{1-10}
Eyeriss & 4.54  & 4.48  & 4.42  & 4.31  & 4.22   & 4.10  & 3.98  & 3.88  & 3.75  &                                                                                    \\ \cline{1-10}
\textbf{TaxoNN}   & 4.84  & 4.78  & 4.70  & 4.65  & 4.49  & 4.31  & 4.15  & 4.13  & 4.04  &                                                                                    \\ \hline
Overhead & 6.5\% & 6.7\% & 6.2\% & 7.9\% & 6.5\%  & 5.2\% & 4.3\% & 6.5\% & 7.7\% & 6.4\%                                                                             \\ \hline
\end{tabular}
}
\vspace{-1em}
\end{table}

Table \ref{summary} shows the neural network accuracy when using TaxoNN with various bitwidths compared to the case of using 32-bit floating-point implementation. Decreasing the bitwidth down to the identified numbers in each configuration set \color{black}  has no considerable impact on the network accuracy. Using a bitwidth lower than the specified one in the configuration sets results in a dramatic accuracy loss as the network can not converge to the desired point. 

\subsection{Hardware Cost}
\label{Hardware cost}
To evaluate the hardware cost of the proposed architecture, we implemented TaxoNN in RTL Verilog and synthesized using the Synopsys Design Compiler with a 45-nm gate library. Table~\ref{area} shows the area cost of the synthesized TaxoNN PE (which supports training) versus the state-of-the-art accelerator, Eyeriss~\cite{eyeriss}, as the baseline architecture (without supporting training). The average area overhead compared to Eyeriss is less than 10\%. The activation functions' derivation unit contributed the most portion of this area overhead and the other units such as the multiplexers had a negligible cost.

Table \ref{power} shows the power consumption of TaxoNN PE compared to that of the Eyeriss architecture (without supporting training) using fixed-point operations. As can be seen, the power consumption is not a concern for TaxoNN due to its pipelines and regular structure. The synthesis results show that the power consumption is, on average, less than 7\% over that of the baseline architecture.
Table \ref{compare-to-FP} summarizes the overall power and area improvement offered by TaxoNN with low-bitwidth operations compared to the full-precision architecture.

\begin{table}[!t]
\centering
\caption{Power and area reduction of TaxoNN compared to the full-precision training implementation}
\vspace{-0.75em}
\label{compare-to-FP}
\renewcommand{\arraystretch}{0.9}
\resizebox{2.5in}{!}
{
\begin{tabular}{|c|c|c|}
\hline
Dataset & \begin{tabular}[c]{@{}c@{}}Power Reduction\end{tabular} & \begin{tabular}[c]{@{}c@{}}Area  Reduction\end{tabular} \\ \hline
MNIST   & 2.1$\times$               & 1.7$\times$                                                      \\ \hline
CIFAR10 & 2.3$\times$                              & 1.8$\times$                                                      \\ \hline
SVHN    & 1.9$\times$                   & 1.5$\times$                                                      \\ 
\hline
\end{tabular}
}
\vspace{-1em}
\end{table}

Moreover, the processing cycles needed for the back-propagation is relatively close to that of feed-forward. Therefore, TaxoNN improves the energy consumption of the training process. These privileges make TaxoNN an appealing accelerator for embedded devices with tight energy constraints.

\section{Conclusions}
In this work, we proposed a light-weight DNN accelerator, called TaxoNN, that supports both inference and training processes. We introduced a novel method to unroll and parallelize the SGD computations. Using this method, we proposed a fine-grained and optimized datapath to perform the matrix operations of SGD. TaxoNN considerably reduces the computation resources required in DNN training by reusing the arithmetic units used in the inference. We evaluated TaxoNN with low-bitwidth operations for each layer. The proposed accelerator offers 1.65$\times$ area and 2.1$\times$ power saving at the cost of\color{black}, on average, 0.97\%  higher misclassification rate compared to the full-precision implementation.


\IEEEtriggeratref{13}
\bibliography{TaxoNN.bib}

\end{document}